# The language of race, ethnicity, and ancestry in human genetic research

Ewan Birney[1,*], Michael Inouye[2,3,4,*], Jennifer Raff[5,6,*], Adam Rutherford[7,*], Aylwyn Scally[8,*]

1. European Molecular Biology Laboratory, European Bioinformatics Institute, Wellcome Genome Campus, Cambridge, United Kingdom
2. Cambridge Baker Systems Genomics Initiative, Department of Public Health and Primary Care, University of Cambridge, Cambridge, UK
3. Cambridge Baker Systems Genomics Initiative, Baker Heart and Diabetes Institute, Melbourne, Victoria, Australia
4. The Alan Turing Institute, London, UK
5. Department of Anthropology, University of Kansas, Lawrence, Kansas, USA
6. Center for Genomics, University of Kansas, Lawrence, Kansas, USA
7. Department of Genetics, Evolution and Environment, University College London, London WC1E 6BT, UK.
8. Department of Genetics, University of Cambridge, Cambridge, United Kingdom

* All authors have contributed equally and are presented alphabetically, their order being entirely arbitrary. Their corresponding emails are: birney@ebi.ac.uk (EB); mi336@medschl.cam.ac.uk (MI); Jennifer.raff@ku.edu (JR); a.rutherford@ucl.ac.uk (AR); aos21@cam.ac.uk (AS)

**Abstract**

The language commonly used in human genetics can inadvertently pose problems for multiple reasons. Terms like 'ancestry', 'ethnicity', and other ways of grouping people can have complex, often poorly understood, or multiple meanings within the various fields of genetics, between different domains of biological sciences and medicine, and between scientists and the general public. Furthermore, some categories in frequently used datasets carry scientifically misleading, outmoded or even racist perspectives derived from the history of science. Here, we discuss examples of problematic lexicon in genetics, and how commonly used statistical practices to control for the non-genetic environment may exacerbate difficulties in our terminology, and therefore understanding. Our intention is to stimulate a much-needed discussion about the language of genetics, to begin a process to clarify existing terminology, and in some cases adopt a new lexicon that both serves scientific insight, and cuts us loose from various aspects of a pernicious past.

**Main text**

The development of genome sequencing has dramatically increased the power of genetics as a tool for biology, anthropology and medicine. Alongside the tangible importance of genetics for medicine and health, its central role in understanding human biology and the ancestral relationships between individuals and groups means that human genetic research can have far-reaching social implications. Therefore, the language that we use to communicate our findings to other researchers, and to the public, is of critical importance.

Accurate terminology is essential in science, but technical language which may enable communication in one context might hinder it in another. Terms which provide a useful shorthand to geneticists may have a subtly or significantly different meaning in other fields, or in common usage (for example, words like mutation, population, heritability, or as discussed below, ethnicity). Some terms may have been inherited from older and less sophisticated (or simply erroneous) literature, or borrowed from other fields.

In the case of human genetics (by which we mean the genetic study of human biology and disease) this often involves terminology inherited from anthropology and population genetics, both of which have evolved significantly in recent years. Both fields now firmly reject longstanding ideas of race as a meaningful biological category and labels which were founded in racist perspectives of the 20[th] and preceding centuries. For example, it is now clear that while humans vary genetically, there is no natural or inherent categorization of human genetic diversity at any global scale. The genetic structures we find are tangled and complex, reflecting our history of worldwide migration, population divergence and admixture, both ancient and recent. While not without consequence for human traits, genetic ancestry and diversity are poorly reflected in the range of anatomical features used in older studies that sought to categorize individuals and groups, features such as pigmentation and cranial or post-cranial morphology. Hence, the racial and ethnographic labels used in such historical studies, and more fundamentally their categorizing approach, have little relevance for understanding human population genetics. Furthermore, the consequences of misunderstanding and misusing these categories, and the terms associated with them, are far from neutral: science, society and politics have always been inherently connected.

As geneticists, we believe that there is an increasing need to address one of the more uncomfortable aspects of this connection, where in the words of American Society of Human Genetics "the invocation of genetics to promote racist ideologies is one of many factors causing racism to persist" [1]. As the source of evidence which some ideologues misrepresent to fit their views of biology and society, the human genetics community has an obligation to revisit and, where possible, change its language to be both more accurate and less confusing or harmful in the public arena. In science, we always aim for precision in our language, but it is not feasible to completely eliminate the possibility of misunderstanding, wilful or otherwise. However, as a community we should strive to foster trust and confidence in our research and its aims, by thinking carefully about the language that we use, as part of an ongoing dialogue with our colleagues and the public. Here, we review how some problematic terms and misconceptions arise from the methods and assumptions commonly used in human genetics, and from the practical challenges of doing human genetics in the modern world. We also discuss potential alternative terminology. Our aim is to stimulate a wider conversation, directed towards finding more precise, and more effective, language and terms that better serve science and avoid historical or politically loaded conventions.

Many of the foundations of physical anthropology, human biology and human classification were developed in service of political ideologies, and the long-term repercussions of this history persist today. One recent example are the attempts by White supremacists in the US to assert 'racial purity' via possession of an allele that bestows lactase persistence (and hence the ability to digest milk post-weaning) and which reached fixation in European populations during the Neolithic. The absurdity of this misapprehension about genetics is highlighted by the fact that this trait does not delineate European ancestry, as equivalent mutations emerged independently and exist at a high frequency in Kazakhs, Ethiopians, Tutsi, KhoeSan, Middle Eastern pastoralists, and amongst many populations where dairy farming has been a significant part of local agricultural practice during their evolutionary history.

Most human geneticists are aware of the problems of imprecise or misused language, but face the difficulty that such language is embedded in many of the methods, tools and data we use. Clinical and anthropological datasets, which can be of enormous utility, often use outdated and scientifically incoherent labels to describe the individuals whose data they include. One example, still frequently used in scientific papers, is 'Caucasian', an 18[th] century term invented to denote pale-skinned northern and western Europeans, or in other archaic connotations a wider range of people based on skull measurements, including west Asians, south Asians, north Africans and Europeans.  In both historical contexts it asserts superiority over other groups and its current usage is therefore laden with scientific, etymological and cultural problems.

Another example in frequent use is "Bantu", which effectively refers to a very broad linguistic grouping comprising hundreds of millions of people in Africa, speaking over 400 distinct languages or dialects. There is some overlap between genetic clusters and Bantu-speaking dialects or languages, but not across the whole group. Furthermore, the word Bantu was used as a catch-all term in apartheid-era South Africa for many different Black African peoples, including groups that were not Bantu-speaking, and had widespread derogatory use in that society. The continued use of these and other similar terms is particularly a problem for longitudinal studies that stretch back decades into the past when such terminologies were current. But even contemporary public health and governmental datasets use census terms which are often arbitrary, outmoded and inconsistent.

Nevertheless, despite their drawbacks and the continuous nature of biological variation in human populations, categories are a necessary part of genetic research and its communication. Genetics affects, to some degree, nearly every human trait or phenotype we can measure, alongside non-genetic factors, often described as

'environmental'. In some cases, we have been able to partly understand the detailed biological mechanisms involved, such as with some alleles of large effect, including those for lactase persistence, or some monogenic disorders like cystic fibrosis. But we lack such understanding for the vast majority of complex traits. As a result, human genetics is an inherently statistical science, one that describes correlations between genomic and phenotypic variation, and attempts to distinguish genetic and environmental effects on phenotypes. The way we group people plays a central role in these analyses, and in many cases, categories enable us to compare and contrast phenotypes and genotypes, and use simpler and more interpretable statistical models, which add to our power of discovery.

Non-genetic factors which influence human phenotypes, also referred to as 'environmental' factors, are diverse and complex, and this complexity is one of the major sources of confusion regarding genetic analyses. One such factor is the cultural and social environment, which includes the racial and ethnic categories to which individuals are assigned (see Box 1 for a discussion of the terms "race" and "ethnicity"). These categories, although socially constructed, can have profound biological effects. For example, by influencing a person's geographical surroundings, their levels of chronic stress, their access to resources, and other aspects of their life history, they may have a major impact on prenatal and childhood development and the expression of human traits.

Thus, the social categories and other groupings that individuals belong to are inescapable components of genetics research. However, within the human genetics community, some aspects of the academic language used to describe groups and subsets of people may foster erroneous beliefs beyond academia about human biology and the nature of these categories. Such descriptions frequently invoke concepts of ancestry and population structure, for reasons we will discuss below. But ancestry itself is often a poorly understood concept, and its relationship to genetic data is not straightforward [2]. There are many implicit assumptions involved in inferring ancestry and population structure, and a similar number of pitfalls when interpreting the output of population genetic clustering analyses and algorithms. For example, the structures found in principal components analysis (PCA) of genetic variation depend strongly on the distribution of genetic ancestry included in the dataset, and is necessarily a sample-specific representation of genetic relationships. Similarly, the clusters identified by widely used methods such as STRUCTURE are often assigned 'ancestry' labels based on the present-day populations within the analysis in which cluster membership happens to be maximised, rather than any explicit inference of ancestral demography. The collection and sampling of genetic data - which often follows existing cultural, anthropological, geographical or political categories - also has a substantial impact, to the extent that some aspects of the clustering reflect sampling strategies rather than any inherent genetic structure [3].

Clusters and labels from population genetic analyses, if used naively, can perpetuate historical or other biases in genetic databases. In more overt cases, these can be commandeered by White supremacists to promote structural racism [4], but there can also be more subtle biases such as the under-representation of non-

European ancestries (for example individuals of African ancestry make up just 2.4% of participants in GWAS studies from 2005 – 2016 [5,6]). This means that many non-European ancestries may be misrepresented in or even excluded from genetic studies, with misleading consequences for the interpretation of their results. At present, alleles associated with disease and other phenotypes are much more likely to have been detected if they are present at intermediate frequencies in European samples than those from Africa and other under-represented regions of the world. This is familiar to human geneticists, but not always communicated effectively outside the field.

More generally, since racial and ethnic group labels are commonly used around the world, particularly in societies with recent origins (e.g., in the Americas) or with post-colonial diversity, their use in genetic research gives the impression that they align with human genetics in some fundamental way. Indeed, it is sometimes assumed, incorrectly, that these labels, self-identified or otherwise assigned, can be used straightforwardly as a proxy for genetic ancestry. This reinforces the commonly-held but simplistic assumption that differences between ethnic groups are substantially due to genetic differences, and are in some sense innate.

For example, researchers working with Indigenous American populations frequently collect genetic samples from only those individuals who have the majority of their genetic ancestry from groups that pre-date European contact (or attempt to mask genomic segments inherited from post-contact admixture in analyses). While this sampling strategy may be useful for understanding certain historical questions about the peopling of the Americas, the exclusion of people with genetic ancestry from more recent admixture events from a cohort labelled "Native American", "Indigenous" or with a more specific population or tribal name implicitly defines these categories in genetic terms. This is both inaccurate and not how Indigenous peoples in the Americas define themselves. By using such labels carelessly, scientists contribute to public misunderstanding of what it means to be an Indigenous person in the Americas [7], and implicitly elevate genetic definitions of "Native American" over tribes' sovereignty in defining their own membership. It can also inadvertently lend support to attempts by individuals not affiliated with Indigenous communities to claim status as "Native American" on the basis of genetic ancestry testing, and apply for certain financial and legal benefits in the United States and Canada [8]. Geneticising the category of "Native American" can also adversely impact medical treatment of community members; assumptions regarding physiology and genetic background may be used by physicians in diagnoses and treatment of community members regardless of whether they accurately apply to specific individuals.

Discussions and close research collaborations between geneticists, biological anthropologists, and Indigenous scholars have led to a greater awareness of these issues in the scientific community. In response, some researchers have made increasing efforts to accurately sample and describe the genetic variation present within Indigenous communities (including in the Americas, ancestry from post-1492 sources), and to more carefully explain and contextualize their use of certain samples for specific questions (such as about historical migrations and population structure). Such research, informed by dialogue with Indigenous communities, strives to use ethnic and community terminologies in the same ways that Indigenous peoples use them, and to reverse the tendency to define (for example) "Native American" in genetic terms.

Yet, despite some recent progress, we believe these problems of language and communication have not concerned the human genetics community as much as they should - indeed it may be that some still do not consider them to be problems at all for our community.

**Human Association Genetics**

A notable example of how the categorization of individuals (and the problem of labelling these categories) is embedded in the methodology of human genetics is the analysis of genetic factors affecting human traits. To discuss this, and what alternative terminology might be preferable, it is necessary to understand the methodological context in more detail. There are two main approaches to discovering such genetic factors. One is linkage studies, in which the variable transmission of genomes and traits is observed in families or larger pedigrees (such as in clinical or population genetic cohorts). These studies often make few assumptions beyond the rules of Mendelian genetic inheritance, which are used to associate genotype with phenotype. The second method, association studies, samples individuals from a population and looks for correlations between genetic and phenotypic measurements. Compared to linkage studies, association studies make more assumptions about how both the underlying population and the non-genetic factors involved are structured. However, due to their applicability to a much wider range of traits, and the increasing capacity to establish large-scale genomic datasets, genome-wide association studies (GWAS) have become a standard method in human genetics for more than 15 years.

Ideally, the main assumption which must be satisfied in an association study of a trait is that non-genetic factors influencing it affect individuals in the cohort randomly with respect to their genotypes. Furthermore, to enhance the power to find genetic effects, minimizing the amount of non-genetic variance is beneficial. If the randomization criteria hold, correlations of genotype to phenotypes can be assigned to the genetic component. However, in real data, the environment is likely to have complex, heterogeneous and ultimately non-random effects on individuals.

As noted above, for many traits the social environment represents an important class of non-genetic factors, ranging from family context to education to many aspects of broader society, including culture, race, and ethnicity. Unsurprisingly, few, if any, groups of humans would fit the assumption of randomness in the environment with respect to genotype. For example, cultural or social differences within a cohort may correlate with differences in ancestry, which will in turn be reflected in genotypic variation. Indeed, regardless of whether the social environment is involved, any correlation between genetic ancestry and the trait will manifest as a signal of association in the study, and thus potentially interpreted as due to genetic (i.e. biological) factors. A number of early genetic association studies in humans were affected by this problem, characterized then as one of significant population structure. Subsequently a PCA-based approach was developed to address this problem, first described in [9], and developed into a standard procedure used today. In this procedure, a subset of the data is selected which is relatively homogeneous with respect to genetic variation. This is done by calculating the principal components (PCs) of genetic variation across samples, and selecting individuals based on their proximity to each other in terms of the top few PCs (sometimes just the first two or three for ease of visualization). Then, in testing for genetic association, a large number of PCs (i.e. the loadings thereof) are used as covariates in the statistical model, with the hope that they will capture variation due to genetic ancestry – and hence also any residual influences of the environment that correlate with it.

Although this procedure is often described as 'removing genetic background effects' or 'correcting for population stratification', neither of these are strictly accurate descriptions, since the aim is not to remove population structure *per se* but primarily to minimise and control for non-genetic effects, for which genetic ancestry differences are treated as a proxy. More problematically, this approach to selecting subsets of individuals from cohorts with ethnicity labels (e.g. "White European", "Han Chinese"), and the implicit assumption of genetic similarity with all members of these categories, has led to such labels being used in a manner very different to their usage in other contexts. Shared culture, as ethnicity implies, does not automatically equate to genetic similarity. Given that genetic association studies are themselves now feeding into genetic prediction analyses using polygenic scores for common traits and diseases (see **Box 2**), this increases the potential for misinterpretation and harm both to individuals and to public understanding. Thus, a lack of clarity about why studies need to account for genetic population structure perpetuates a biological essentialism implicit in the labels used, and gives the impression that the goal of human genetics research is to characterise the genetics of different ethnicities rather than to better understand human biology by controlling for non-genetic effects. It must be clear to all that our goal is the latter.

It should be acknowledged that not all these labels are without relevance, in that for many there may indeed be social environments which can meaningfully be described in such terms and which may affect the genetics of some traits. Nevertheless, framing social environments in purely genetic terms may seem to imply a genetic explanation, rather than societal, cultural and historical explanations, for the existence of these categories and the differences between them. Moreover, some of the labels used are simply archaic and discredited terms from earlier eras of anthropology and human genetics, and therefore inappropriate as descriptors of human groupings today.

**Proposed Language**

We argue that a more critical understanding by geneticists of the societal impact of their work and how it is communicated is necessary in order to minimise or prevent misunderstanding. A practical implication of this is the need for new concepts, terminology and language to enable scientists to communicate accurately and appropriately both within their field and to other scientific and lay audiences.

Humans are social creatures, and it is no surprise that social interactions have profound effects on many human phenotypes, sometimes outweighing the effect of an individual's own genetic make-up. Although often the goal in studying genomes is to find genetic causes of phenotypic variation, in practice this tends to minimize other sources of variation outside of genetics, including social variation. In particular, we call for an approach to describing methods and results which recognizes the large effects of non-genetic factors--particularly social, cultural, and economic--on human traits and uses language that reflects the entangled nature of culture and genetics in human life. There are other fields - anthropology, psychology and sociology - with extensive histories of describing the impacts of interpersonal behaviours in family, educational and cultural settings, and we advocate for a deeper engagement with researchers from these fields - ideally through sustained research collaboration - in helping shape the terminology we use and even the design of our studies.

We highlight in **Table 1** some terms and language constructs used in human genetics which we believe are particularly prone to being misleading and unhelpful for scientific communication. We have suggested alternative forms of language to use in each case, generally aiming for technical accuracy over concision. While there is a natural tendency to avoid jargon and unwieldy language when communicating ideas, the drawback can be a lack of precision, and a false familiarity when terms or labels from common usage or other fields are used. The risks of misunderstanding here are such that the brevity of our language is less important than its clarity and accuracy regarding the methods, categories and concepts involved and their relationship to human populations and individuals.  Some of these suggestions may meet with disagreement; we present them partly to stimulate discussion of these and other terms, and in the hope that this will lead to better and more accurate language conventions and less misunderstanding, particularly outside of human genetics.

**Table 1**

| Term or concept | Issues | Alternative terminology or approach |
|---|---|---|
| "Caucasian" | Use of old terms associated with racist and pseudo-scientific classification of humans. | Avoid usage where possible. When usage is unavoidable, e.g. the term is a label in the dataset used, explicitly note this, emphasise that it has no scientific validity, and use quotes when referring to the label in subsequent analyses. |
| "Bantu" "Eskimo-Aleut" | Misapplication of linguistic, ethnic, or other historical terms describing human social or cultural structures. | In describing a group, cohort or data used, where possible use cultural and other terms defined or accepted by the group themselves. Avoid ethno-linguistic terms that more properly refer to wider or more diverse groups than are represented in the study cohort.<br><br>For example, 'Bantu' has coherence as a (broad) linguistic group (and therefore should normally only be used with the suffix '-speaking'), but note how broadly "Bantu-speaking" applies across the African continent (as discussed in the text).<br><br>In particular, avoid using linguistic or cultural terms to differentiate groups identified on the basis of genetic information alone, such as PCA-selected groups (see below). |
| "Genetic background" | This term is commonly used to describe genetic differences between laboratory or domesticated animals such as mice, particularly differences that are not explicitly specified. In a human context it could refer to the shared genetic variation that distinguishes one group of individuals from another. However, the very great differences between human populations and laboratory animal strains or domesticated breeds makes this a highly misleading usage. Often it carries the implication that trait differences are purely or primarily genetic in origin. In many situations, this is confused with social stratification coupled | Implicit or non-specified genetic differences between individuals or groups are better referred to as differences in genetic ancestry. It should be understood that human genetic ancestry is complex, and no human population or group, even one defined by genetic clustering, constitutes a consistent 'background' against which genetic or phenotypic variation can be compared. In discussing trait differences between cohorts, consider that studies are often unable to explore whether the observed differences between groups were due to genetic or cultural effects due to the confounding of cultural components and genetic structure. |
| | with non-randomness in human studies. | |
| "Ethnicity" | Ethnicity, and labels for specific ethnicities (e.g. "Hispanic", "White Irish", "Native American") are complex and loaded terms with specific meanings in different contexts. (In human genetic data they typically correspond to categories in censuses and surveys). Often they are misused as proxy terms for genetic ancestry, and hence indirectly to represent different social environments. For example, in some contexts in the United States, people considered "Hispanic" may have highly varying proportions of ancestry from pre-contact Indigenous populations, settlers from European countries, enslaved peoples from Africa, and recent immigrants from across the globe.

As discussed above, "Native American" is a social and political category within the United States (within Canada the term "First Nations" is more often used); it may or may not reflect ancestry from pre-contact Indigenous populations. | Avoid using these terms without an understanding of their context and how to appropriately apply them. When used, make such context explicit (See **Box 1**).

Where labels for certain ethnicities are already used within a dataset, e.g. on survey forms or other previously-gathered data, explicitly note the survey or data they derive from and the fact that they are not assignments of genetic ancestry.

Where labels are self-identified, be explicit about this in category names, e.g. 'self-identified Hispanic',

Use broad social terms if commenting on social structures: "the heterogeneity in social environment between the subsets is large, and makes downstream comparisons complex"

Wherever possible, also use a name by which the group identifies itself: "member of [X tribe or nation]" |
| Cultural, historical, linguistic or ethnic terms (e.g. "European/Han Chinese/Japanese") used to label a group of individuals identified on the basis of genetic data, e.g. by PCA clustering. | Groupings defined by genetic variation do not necessarily correspond with cultural, linguistic or ethnic categories, and usually exclude many individuals to whom such terms apply, as well as including individuals to whom they do not.

For example, even within a cohort many 'European' or 'Han Chinese' individuals may be excluded from a particular cluster in PCA space, and such terms, understood as ethnic or cultural identities, potentially represent a much greater range of genetic ancestry. Thus, it is misleading to attach them to the results of subsequent analyses based on such clustering. | Include appropriate qualifiers, such as 'predominantly', 'culture', 'associated', or 'PCA-selected' depending on the method.

e.g., "The European-associated PCA cluster, which aims to minimise variation in non-genetic factors and genetic factors".

If referring to the technique in general of taking a subset of individuals to minimise cultural components, be explicit "predominantly European-American culture subset of cohorts, often selected via genetic principal components" or "predominantly Japanese culture subset";

For repeated use of terms, aim to always qualify cultural terms, eg, "PCA-selected European subset" |

| | | |
|---|---|---|
| | | rather than "the European subset" with no qualifiers |
| "Population structure" | Population structure in humans is ubiquitous and complex, and manifests as genetic differences between individuals and groups. *A priori* these differences cannot be assumed to have a significant effect on a given trait, compared to effects of the social environment which may correlate with population structure. Nor can it be assumed that such genetic effects that *are* associated with population structure will necessarily act in the same direction as observed phenotypic differences. | If the study focuses on allele frequency shifts between groups of people, highlight the specific allele frequency shifts of interest. If the study assesses between-cohort phenotypic differences, consider the likely confounding between population structure and social environment, and hence avoid implicitly or explicitly ascribing phenotypic differences to population structure or differences in genetic ancestry. |

**Box 1: The terms 'race' and 'ethnicity'**

Race and ethnicity are important aspects of the social environment for many human traits, but these words have (and have had) a variety of meanings in different contexts and different parts of the world. Their use must therefore be approached with care, whether in the context of clinical or census data where participants have been asked to self-identify in these categories, or in more general discussion. 'Race' is particularly problematic, and its historical and political connotations, along with the fact that it is not a meaningful descriptor of genetic variation, have led many human geneticists to avoid it altogether. Indeed, in usage outside the United States, 'race' is less consistently understood and 'ethnicity' is often viewed as a less contentious way of referring collectively to those elements of an individual's identity and biology that are inherited through ancestry and culture.

By contrast, in the United States, and within anthropological genetics (a subfield of biological anthropology), race and ethnicity have separate and distinct meanings; the former is a socially constructed category that takes into account physical characteristics, and the latter is a explicitly category of cultural self-identification. This usage (which has itself changed considerably over the years since the United States government began collecting census information) reflects a complex history of colonialism, politics and attitudes to race [10].

Since we are primarily addressing colleagues in genetics, and because we feel that 'race' is particularly liable to misinterpretation in a genetic context, we have leaned towards a broader meaning of ethnicity. But the variable meanings of these words must be considered when communicating genetic research, even when these ideas themselves are not its focus, because they are central to how people interpret differences between human groups and individuals. They will also continue to evolve, and reaching a consensus across disciplines and cultures will continue to require extensive discussion and debate. But this is not entirely a drawback, as such discussions can bring us to a deeper understanding of each other's fields, cultural backgrounds, and our own assumptions.

**Box 2: Polygenic risk scores and ancestry**

Relatively recently it has become possible - and increasingly common - to use large-scale genome-wide data, individually and/or in combination with GWAS summary statistics, to estimate an individual's genetic predisposition for a particular complex trait or disease. This predisposition is, at its simplest, a single number or score based on the relative contribution (weighting) of multiple genetic variants to the phenotype. In contrast to monogenic variation, the use of genome-wide data covering a large number of genetic loci means that these scores are polygenic and therefore they are termed 'polygenic scores' (and frequently 'polygenic risk scores' if they relate to discrete phenotypes such as disease status).

Since a polygenic score is based on genetic variants across the genome, the polygenic score's distribution within a population depends exquisitely on the allele frequencies and linkage disequilibrium in a population (and the extent to which they differ from the original cohort(s) used to develop the polygenic score). This means it is difficult to compare and interpret scores from individuals of different populations, particularly those with large differences in ancestry or culture. If one were to compare a polygenic score from an individual of Dutch ancestry to that of an individual of Japanese ancestry, they may have very different scores, but their phenotypes may well be very similar. Therefore, polygenic scores carry with them a double-edged sword: they can have statistical meaning (and potential clinical utility), but if used carelessly without respect for the population history and social environment, conclusions based on polygenic scores can be invalid and have the potential to exacerbate both health and socioeconomic inequalities.

The classification of individuals into discrete populations or ancestries is itself a compromise of necessity. *All* individuals are admixed, and all human genomes comprise multiple ancestries. With genomics and genome sequencing we are able to infer and model the ancestry variation in different parts of chromosomes, including the many ancestries shared widely with individuals in different populations around the world, (e.g. those from extinct types of human, such as Neanderthals or Denisovans), and others found only in a few present-day populations.

Given the reality of the admixed individual, the approximation of the polygenic score becomes more relevant, i.e., each individual genome has multiple ancestries and any current polygenic score sums over all of these. This is best exemplified by individuals of recent diverse admixture (e.g., mother and father from ancestries of different continents), for whom the utility of current polygenic scores is unclear. Yet, the approximation which gets us closer to the truth is that each locus (or haplotype) within an individual has its own ancestral history and its own effects on downstream phenotypes. As more human genomes are sequenced and individual phenotypes quantified, our inferences of the family tree which links humanity and the ancestral context of each phenotype correlation of each haplotype will be improved. With these data, we will better understand the ancestries and histories which comprise each of our genomes and which drive the molecular and cellular processes that interface our (social) environments. Alongside these scientific advances, we foresee an empirical basis and opportunity to correct the commonplace but imprecise language in human genetics which unfortunately results in public misunderstanding.

**Acknowledgements.**

We thank Eimear Kenny, Tuuli Lappalainen, Agustin Fuentes and Carina Schlebusch for comments and contributions to early versions of this article.